\documentclass[11pt]{iopart}
\usepackage{iopams} 
\usepackage{setstack,amsthm,epsfig,graphicx,cite,color,multicol,subfigure}
\usepackage{bm,array,calc,url}
\usepackage{hyperref}

\newcommand{\be}{\begin{equation}}
\newcommand{\ee}{\end{equation}}
\newcommand{\bea}{\begin{eqnarray}}
\newcommand{\eea}{\end{eqnarray}}
\newcommand{\nn}{\nonumber}

\def\bra#1{|#1\rangle}
\newcommand{\ben}{\begin{eqnarray*}}
\newcommand{\een}{\end{eqnarray*}}

\def\i{{\rm i}}
\def\e{{\rm e}}
\def\qpch#1{(#1;q)_\infty}
\def\zetac{\zeta_{\rm c}}

\def\g{\gamma}
\def\b{\beta}
\def\d{\delta}
\def\a{\alpha}
\def\e{\varepsilon}

\def\nn{\nonumber\\}

\def\2t#1#2{\langle\tau_{#1}\tau_{#2}\rangle}

\def\nn{\nonumber\\}
\def\tt{\tilde{\theta}}
\def\fr#1{(\ref{#1})}

\def\i{{\rm i}}
\def\d{{\rm d}}
\def\e{{\rm e}}
\def\dps{\displaystyle}

\eqnobysec
\begin{document}

\title[]{Relaxation rate of the reverse biased asymmetric exclusion process}
\author{$^1$Jan de Gier, $^2$Caley Finn and $^3$Mark Sorrell}
\address{Department of Mathematics and Statistics, The University of Melbourne, 3010 VIC, Australia}
\eads{${}^1$jdgier@unimelb.edu.au, ${}^2$c.finn3@pgrad.unimelb.edu.au, ${}^3$msorrell@unimelb.edu.au}

\begin{abstract}
We compute the exact relaxation rate of the partially asymmetric exclusion process with open boundaries, with boundary rates opposing the preferred direction of flow in the bulk.  This reverse bias introduces a length scale in the system, at which we find a crossover between exponential and algebraic relaxation on the coexistence line. Our results follow from a careful analysis of the Bethe ansatz root structure.
\end{abstract}

\maketitle



\section{Introduction}
The partially asymmetric exclusion process (PASEP) \cite{ASEP1,ASEP2} is one of the most thoroughly studied models of non-equilibrium
statistical mechanics \cite{Derrida98,Schuetz00,GolinelliMallick06,BEreview}. It is a microscopic model of a driven system \cite{schmittmann} describing the asymmetric diffusion of hard-core particles along a one-dimensional chain. In this paper we will consider a finite chain with $L$ sites. At late times the PASEP exhibits a relaxation towards a non-equilibrium stationary state. In the presence of two boundaries at which particles are injected and extracted with given rates, the bulk behaviour at stationarity is strongly dependent on the injection and extraction rates. The corresponding stationary phase diagram as well as various physical quantities have been determined by exact methods \cite{Derrida98,Schuetz00,GolinelliMallick06,DEHP,gunter,sandow,EsslerR95,PASEPstat1,PASEPstat2,BEreview}.

More recently exact dynamical properties such as relaxation rates have been derived. On an infinite lattice where random matrix techniques can be used, considerable progress has been made, see e.g. \cite{gunter97,johansson,praehoferS02a,TracyWidom08}.
For the PASEP on a finite ring, where particle number is conserved, relaxation rates were obtained by means of the Bethe ansatz some time ago \cite{dhar,BAring1a,BAring1b,BAring2}. The dynamical phase diagram of the PASEP with open boundaries was studied in \cite{GierE05,GierE06,dGE08}, again using Bethe ansatz methods. 

In the present work we extend some of these results to the case of reverse bias, where the boundary parameters strongly oppose the bias in the bulk hopping rates. We will in particular focus on the coexistence line where the effects are most pronounced.

\subsection{The partially asymmetric exclusion process}
\begin{figure}[ht]
\centerline{
\begin{picture}(320,80)
\put(0,0){\epsfig{width=0.7\textwidth,file=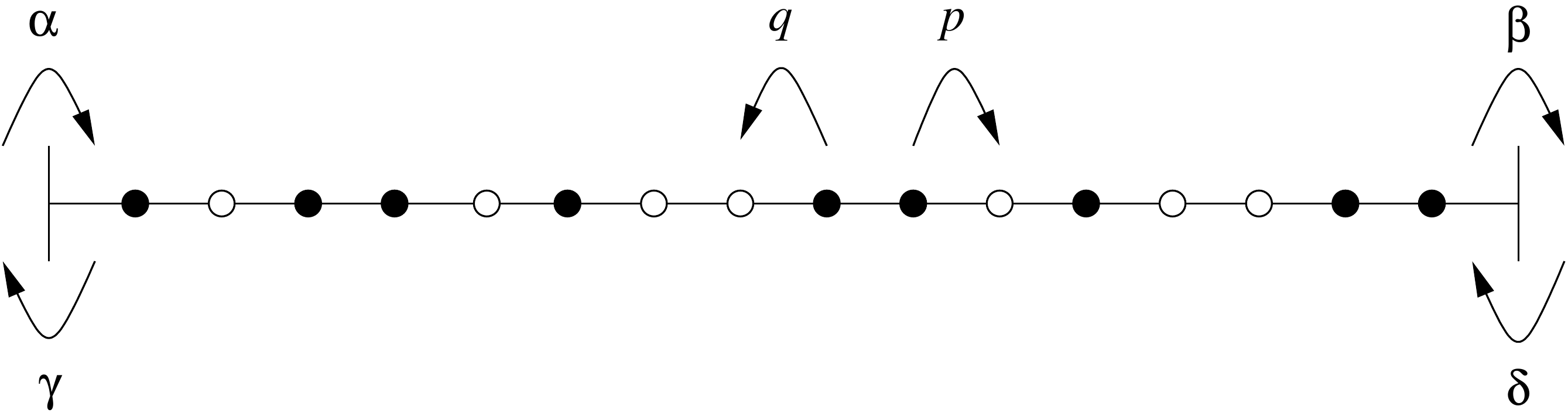}}
\end{picture}}
\caption{Transition rates for the partially asymmetric exclusion process.}
\label{fig:paseprules}
\end{figure}

We now turn to a description of the dynamical rules defining the PASEP on a one dimensional lattice with $L$ sites, see Figure~\ref{fig:paseprules}. At any given time
$t$ each site is either occupied by a particle or empty. The system is then updated as follows. 

Particles attempt to hop one site to the right with rate $p$ and one site to the left with rate $q$. The hop is prohibited if the neighbouring site is occupied, and at sites $1$ and $L$ these rules are modified in the following way. If site $i=1$ is empty, a particle may enter the system with rate $\alpha$. If, on the other hand, this site is occupied, the particle will leave the system with rate $\gamma$ or attempt to hop to the right with rate $p$. Similarly, at $i=L$ particles are injected and extracted with rates $\delta$ and $\beta$ respectively, and attempt to hop to the left with rate $q$.

It is customary to associate a Boolean variable $\tau_i$ with every
site, indicating whether a particle is present ($\tau_i=1$) or not
($\tau_i=0$) at site $i$. Let $\bra0$ and $\bra1$ denote the standard
basis vectors in $\mathbb{C}^2$. A state of the system 
at time $t$ is then characterised by the probability distribution
\be
\bra{P(t)} = \sum_{\bm \tau} P(\bm{\tau}|t) \bra{\bm{\tau}},
\ee
where
\be
\bra{\bm{\tau}} = \bra{\tau_1,\ldots,\tau_L} = \bigotimes_{i=1}^{L} \bra{\tau_i}.
\ee
The time evolution of $\bra{P(t)}$ is governed by the aforementioned
rules, which gives rise to the master equation
\bea
\frac{{\rm d}}{{\rm d} t} \bra{P(t)} &=& M \bra{P(t)},
\label{eq:Markov}
\eea
where the PASEP transition matrix $M$ consists of two-body
interactions only and is given by 
\be
M = \sum_{k=1}^{L-1} I^{(k-1)}\otimes M_k\otimes I^{(L-k-1)} + m_1 \otimes I^{(L-1)}+  I^{(L-1)}\otimes m_L.
\label{eq:TransitionMatrix}
\ee
Here $I^{(k)}$ is the identity matrix on the $k$-fold tensor product
of $\mathbb{C}^2$ and, for $1\leq k \leq L-1$, $M_k: \mathbb{C}^2\otimes \mathbb{C}^2
\rightarrow  \mathbb{C}^2\otimes \mathbb{C}^2$ is given by
\bea
M_k = \left(\begin{array}{@{}cccc@{}}
0 & 0 & 0 & 0\\
0 & -q & p & 0 \\
0 & q & -p & 0 \\
0 & 0 & 0 & 0
\end{array}\right).
\eea
The terms involving $m_1$ and $m_L$ describe injection
(extraction) of particles with rates $\a$ and $\delta$ ($\g$ and $\b$) at
sites $1$ and $L$ respectively. Their explicit forms are
\be
m_1=\left(\begin{array}{@{}cc@{}}-\a&\g \cr \a &-\g\cr\end{array}\right),\qquad 
m_L= \left(\begin{array}{@{}cc@{}}-\delta&\b\cr \delta&-\b\cr \end{array}\right).
\label{h1l}
\ee

The transition matrix $M$ has a unique stationary state
corresponding to the eigenvalue zero. For positive rates, all
other eigenvalues of $M$ have non-positive real parts. The late time
behaviour of the PASEP is dominated by the eigenstates of $M$ with the
largest real parts of the corresponding eigenvalues.

In the next sections we set $p=1$ without loss of generality and determine the eigenvalue of $M$ with the largest non-zero
real part using the Bethe ansatz. The latter reduces the problem of
determining the spectrum of $M$ to solving a system of coupled
polynomial equations of degree $3L-1$. Using these equations, the
spectrum of $M$ can be studied numerically for very large $L$, and, as
we will show, analytic results can be obtained in the limit for large $L$. 

\subsection{Reverse bias}
The reverse bias regime corresponds to the boundary parameters opposing the direction of flow in the bulk. In this paper we achieve this by considering the regime
\be
   q < 1, \quad \alpha, \beta \ll \gamma, \delta.
\label{eq:rb}
\ee
With $q<1$ the bias in the bulk is for diffusion left to right, but setting $\alpha,\beta \ll \gamma,\delta$ strongly
favours particles entering from the right and exiting on the left. Traditionally the reverse bias regime was considered only to exist for the case $\alpha=\beta=0$, see e.g. \cite{PASEPstat2}\footnote{Note that the authors of \cite{PASEPstat2} consider $q>1$ and $\gamma=\delta=0$ but we can simply translate their results to our notation using left-right symmetry.}

\section{Stationary phase diagram}
In this section we briefly review some known facts about the stationary phase diagram. 
It will be convenient to use the following standard combinations of parameters \cite{sandow},
\be
\kappa^{\pm}_{\alpha,\gamma} = \frac{1}{2\alpha} \left(
v_{\alpha,\gamma} \pm \sqrt{v_{\alpha,\gamma}^2 +4\alpha\gamma}\right),\qquad 
v_{\alpha,\gamma} = 1-q-\alpha+\gamma.
\ee
In order to ease notations we will use the following abbreviations, 
\be
a=\kappa^+_{\alpha,\gamma},\quad b=\kappa^+_{\beta,\delta},\quad c=\kappa^-_{\alpha,\gamma},\quad d=\kappa^-_{\beta,\delta}.
\label{eq:ab}
\ee
The parameters $a,b,c,d$ are standard notation for the parameters appearing in the Askey-Wilson polynomials $P_n(x;a,b,c,d|q)$ used for the PASEP sationary state \cite{UchiSW}.

\subsection{Forward bias}
\begin{figure}[ht]
\centerline{\includegraphics[width=200pt]{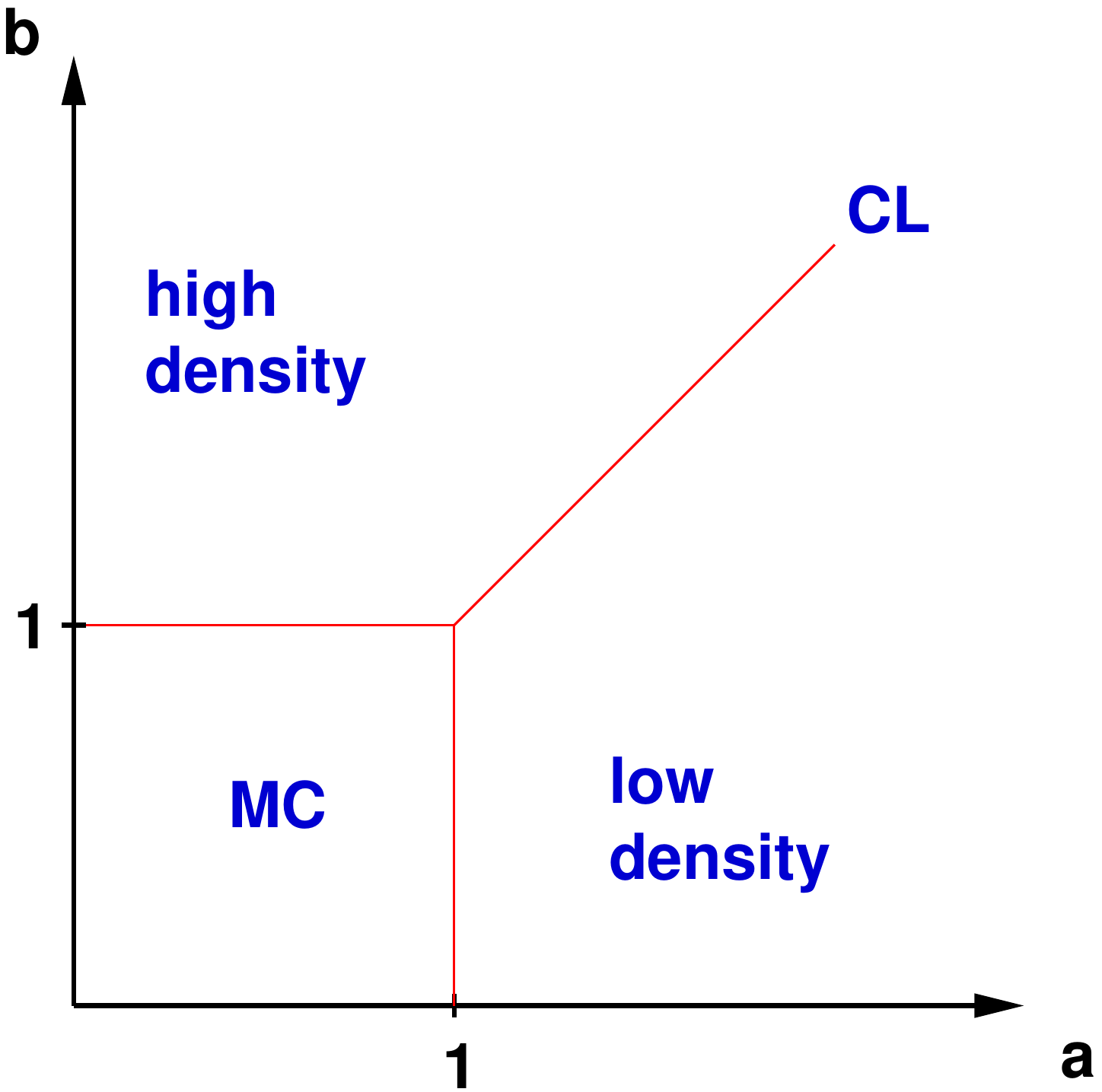}}
\caption{Stationary state phase diagram of the PASEP.
The high and low density phases are separated by the
coexistence line (CL). The maximum current phase (MC) occurs at small
values of the parameters $a$ and $b$ defined in (\ref{eq:ab}). } 
\label{fig:statPD}
\end{figure}

The phase diagram of the PASEP at stationarity was found by Sandow \cite{sandow} and is depicted in Figure~\ref{fig:statPD}. The standard phase diagram is understood to depend only on the parameters $a$ and $b$ defined in (\ref{eq:ab}) rather than $p,q,\alpha,\beta,\gamma,\delta$ separately, see e.g. the recent review \cite{BEreview}. However, we will show below that the picture may be more nuanced in the regime where $\alpha, \beta \ll \gamma, \delta$.

Many quantities of interest, such as density profiles and currents, have been calculated for particular limits of the PASEP. For the most general case these have been computed by \cite{UchiSW,UchiW}.

\subsection{Reverse bias}
Much less is known for the case of reverse bias.
The stationary state normalisation and current for $\alpha=\beta=0$ have been computed by \cite{PASEPstat2}. It is found that the current decays exponentially as
\be
J = (1-q^{-1}) \left(\frac{\gamma\delta}{(1-q+\gamma)(1-q+\delta)} \right)^{1/2} q^{L/2-1/4}.
\ee

It was argued in \cite{PASEPstat2} that non-zero values of $\alpha$ and $\beta$ would allow for a right flowing current of particles to be sustained, thus destroying the reverse bias phase. However, in the following we will see that when $\alpha$ and $\beta$ are $\mathcal{O}(q^m)$ for some $m>0$, the system still feels the effects of a reverse bias in its relaxation rates. It would be interesting to know whether such effects survive in the stationary state.

\section{Relaxation rates -- approach to stationarity}

The slowest relaxation towards the stationary state at asymptotically late times $t$ is given by $\e^{\mathcal{E}_1 t}$, where $\mathcal{E}_1$ is the eigenvalue of the transition matrix $M$ with the largest non-zero real part.

It is well-known that the PASEP can be mapped onto the spin-1/2 anisotropic Heisenberg chain with general (integrable) open boundary conditions \cite{sandow,EsslerR95}. By building on recent progress in applying
the Bethe ansatz to the diagonalisation of the Hamiltonian of the latter problem \cite{Cao03,Nepo02,YangZhang}, the Bethe ansatz equations for the PASEP with the most general open boundary conditions can be obtained. Recently these equations have also been derived directly using a (matrix) coordinate Bethe anstaz in the PASEP \cite{Simon09,CrampeRS}. The Bethe ansatz equations diagonalise the PASEP transition matrix. By analysing this set of equations for large, finite $L$ the eigenvalue of the transition matrix with the largest non-zero real part has been obtained in the forward bias regime \cite{GierE05,GierE06,dGE08}. In this regime, particles diffuse predominantly from left to right and particle injection and extraction occurs mainly at sites $1$ and $L$ respectively. We will briefly review these results.

\subsection{Bethe ansatz equations}

As was shown in \cite{GierE05,GierE06,dGE08}, all eigenvalues ${\cal E}$ of $M$ other than ${\cal E}=0$ can be expressed in terms of the roots $z_j$ of a set of $L-1$ non-linear algebraic equations as 
\bea
{\mathcal{E}}= -\mathcal{E}_0-\sum_{j=1}^{L-1}\varepsilon(z_j),
\label{eq:pasep_en}
\eea
where
\be
\mathcal{E}_0 = \alpha+\beta+\gamma+\delta = (1-q)\left(\frac{1-ac}{(1+a)(1+c)}
+\frac{1-bd}{(1+b)(1+d)}\right),
\label{eq:e0}
\ee
and the ``bare energy'' $\varepsilon(z)$ is
\begin{equation}
   \varepsilon(z) = \frac{(q-1)^2 z}{(1-z)(qz-1)} =(1-q)\left(\frac{1}{z-1}-\frac{1}{qz-1}\right). 
\label{eq:epsdef}
\end{equation}
The complex roots $z_j$ satisfy the Bethe ansatz equations
\bea
\left[\frac{qz_j-1}{1-z_j}\right]^{2L} K(z_j) =\prod_{l\neq j}^{L-1}
\frac{qz_j-z_l}{z_j-qz_l} \frac{q^2z_jz_l-1} {z_jz_l-1},\
j=1\ldots L-1.\nn
\label{eq:pasep_eq}
\eea
Here $K(z)$ is defined by
\be
  K(z) =
\frac{(z+a)
  (z+c)}{(qaz+1)
  (qcz+1)} \frac{(z+b)
  (z+d)}{(qbz+1)
  (qdz+1)},
\ee
and $a,b,c,d$ are given in \eref{eq:ab}.

\subsubsection{Analysis in the forward bias regime}
The analysis of the Bethe equations for the forward bias regime proceeds along the following lines. One first defines the counting function,
\be
\i Y_L(z) = g_{\rm}(z) + \frac{1}{L} g_{\rm b}(z)
+ \frac{1}{L} \sum_{l=1}^{L-1} K(z_l,z),
\label{eq:logtasepBAE}
\ee
where
\bea
g_{\rm }(z) &=& \ln \left( \frac{z(1-qz)^2}{(z-1)^2}\right), \nonumber\\
g_{\rm b}(z) &=& \ln \left(\frac{z(1-q^2z^2)}{1-z^2}\right) +
\ln\left(\frac {z+a}{1+qaz}\frac{1+c/z}{1+qcz}\right) \nonumber\\
&& {}+
\ln\left(\frac {z+b}{1+qbz}\frac{1+d/z}{1+qdz}\right),
\label{eq:kernelDefFB}\\
K(w,z) &=& -\ln(w) -\ln \left( \frac{1-q z/w}{1-qw/z} \frac{1-q^2zw}{1-w z}\right).
\nonumber
\eea
We note that the choice of branch cuts in this definition of $K(w,z)$ differs from \cite{dGE08}, but is numerically stable for large $L$. A further discussion on the choice of branch cuts in $K$ is given in Section~\ref{sec:revasymp}.

Using the counting function, the Bethe ansatz equations \fr{eq:pasep_eq} can be cast in logarithmic form as
\be
Y_L(z_j) = \frac{2\pi}{L} I_j\ ,\qquad j=1,\ldots,L-1,
\label{eq:Z=I}
\ee
where $I_j$ are integers. Each set of integers $\{I_j|\; j=1,\ldots, L-1\}$ in (\ref{eq:Z=I}) specifies
a particular (excited) eigenstate of the transition matrix.

In \cite{dGE08}, the logarithmic Bethe equations were solved numerically to find the structure of the root distribution (an example of such a solution will be seen in the next section). By comparing the resulting eigenvalues with brute force diagonalisation of the transition matrix, the integers $I_j$ corresponding to the first excited state can be found for small values of $L$. The integers were found to be
\be
I_j = -L/2+j\quad {\rm for}\quad j=1,\ldots,L-1.
\label{eq:Idef}
\ee
Assuming \eref{eq:Idef} holds for all $L$, \fr{eq:Z=I} can be turned into an integro-differential equation for the
counting function $Y_L(z)$, which can be solved asymptotically in the limit of large $L$.  Exact asymptotic
expressions were obtained for the first excited eigenvalue in the forward bias regime in all areas of the
stationary phase diagram except the maximal current phase. We recall here the expression for the coexistence
line:
\be
\mathcal{E}_1 = \frac{1-q}{L^2} \frac{\pi^2}{(a^{-1}-a)} + \mathcal{O}(L^{-3}).
\label{eq:E_CL}
\ee
%

\subsection{Dynamical phase diagram}

\begin{figure}[ht]
\centerline{\includegraphics[width=250pt]{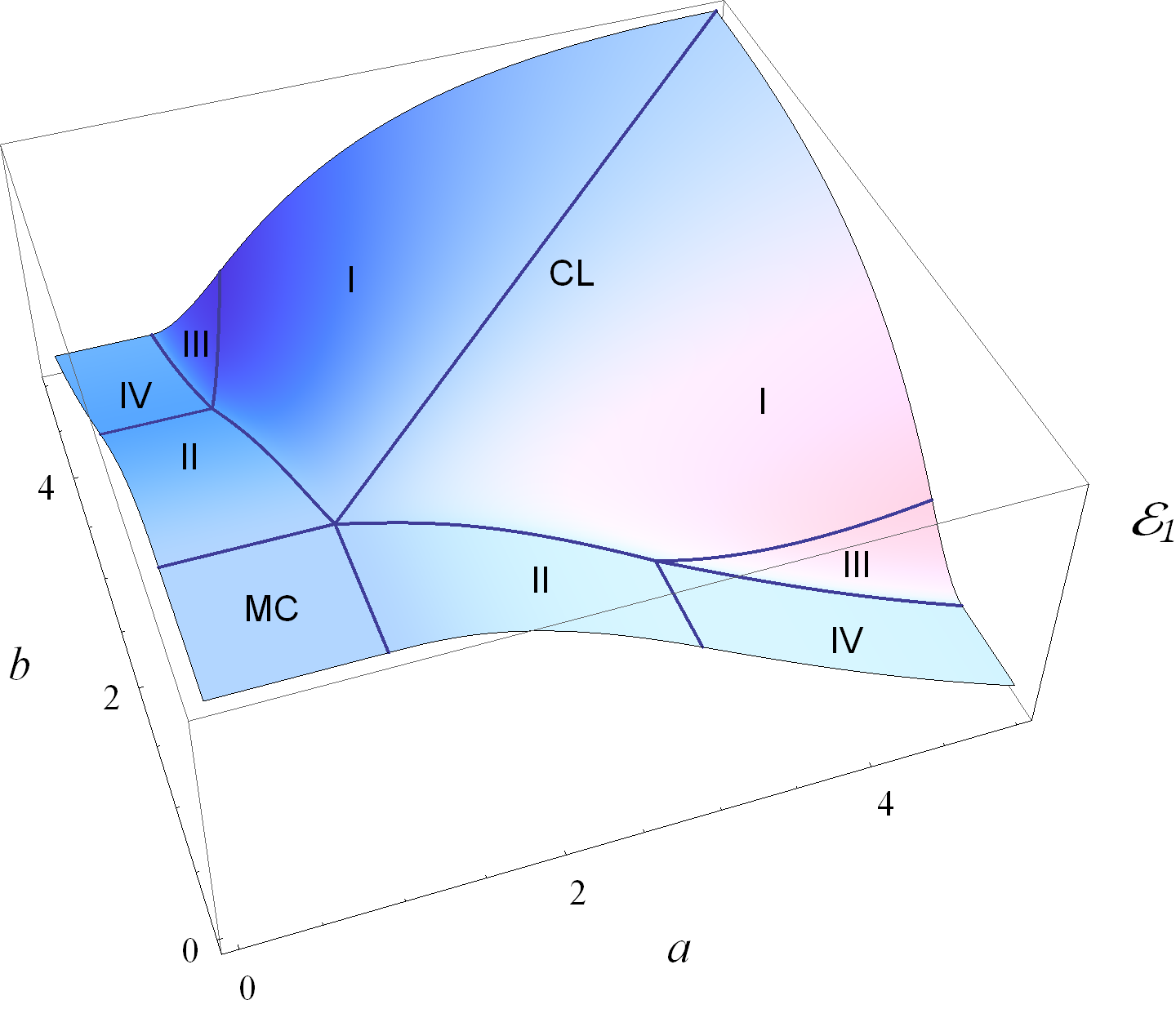}}
\caption{Dynamical phase diagram of the PASEP in the forward bias
regime determined by the lowest excitation $\mathcal{E}_1$. The
horizontal axes are the boundary parameters $a$ and $b$ \fr{eq:ab} and
the vertical axis is the lowest relaxation rate. The latter goes to
zero for large systems on the coexistence line (CL) and in the maximum
current phase (MC). The curves and lines correspond to various
crossovers in the low and high density phase, across which
the relaxation rate changes non-analytically. }  
\label{fig:phasediagram}
\end{figure}

The dynamical phase diagram for the PASEP resulting from an analysis in the regime $q<1$ is shown in Figure~\ref{fig:phasediagram}. It exhibits the same subdivision found from the analysis of the current in the stationary state: low and high density phases ($a>1$ and $b>1$) with dynamic exponent $z=0$, the coexistence line ($a=b>1$) with diffusive ($z=2$) behaviour and the maximum current phase ($a,b<1$) with KPZ ($z=3/2$) universality. Furthermore, the finite-size scaling of the lowest excited state energy of the transition matrix suggests the sub-division of both low and high-density phases into four regions respectively. These regions are characterized by different functional forms of the relaxation rates at asymptotically late times \cite{dGE08}, see also \cite{ProemeBE}.

\section{Bethe ansatz in the reverse bias regime}
\label{se:BAanalysis}

The reverse bias regime corresponds to taking parameters $a$, $b$ large, and $c$, $d$ towards $-1$ (note that
$-1 < c,d \le 0$).  This change, and the resulting changes in the structure of the root distribution, requires
a reformulation of the counting function, which will be discussed in Section~\ref{sec:revasymp} and
\ref{ap:numsols}.  But first we describe the changes to the root structure.  In the following we address
solutions on the coexistence line, $a = b$, and to simplify computations further also set $c = d$.
\begin{figure}[b]
\begin{center}
   \includegraphics{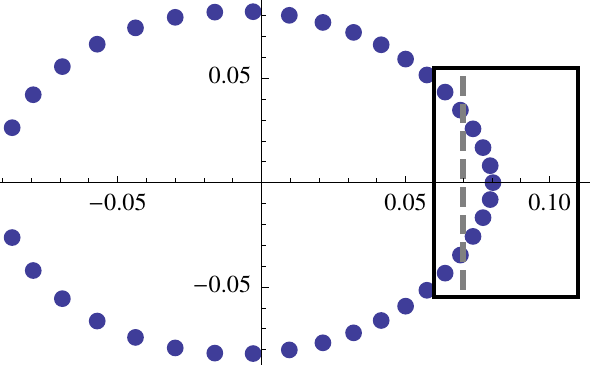}
   \caption{Root distribution in the forward bias regime for $L = 40$, with parameters $a = b = 11$, $q = 0.3$
            and $c = -0.07$.  The dashed line marks the position of $-c$ and the region in the box is expanded
            in Figure \ref{fig:roots_forward} below.}
   \label{fig:roots_forward_full}
\end{center}
\end{figure}

\begin{figure}[h!]
\centering
\subfigure[$c = -0.07$]{
  \includegraphics{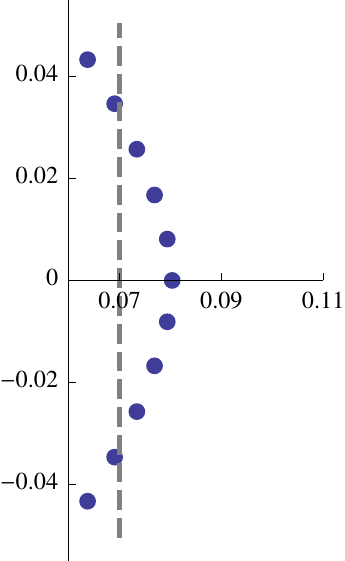}
   \label{fig:roots_forward}
 }
 \subfigure[$c = -0.088$]{
  \includegraphics{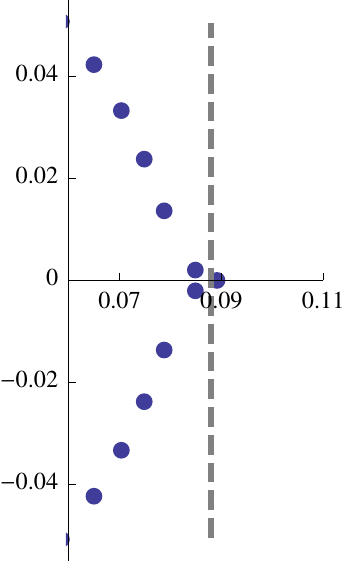}
   \label{fig:roots_interm}
 }
 \subfigure[$c = -0.1$]{
  \includegraphics{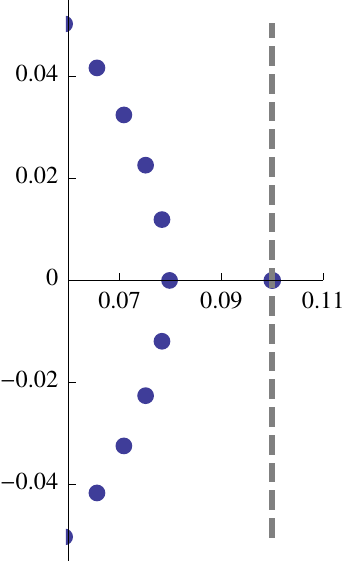}
   \label{fig:roots_reverse}
 }

\caption{Leading part of the root distribution for $L=40$, $a = b = 11$, $q = 0.3$, with $c = -0.07$ (forward bias),
$c = -0.088$ (intermediate stage), and $c = -0.1$ (reverse bias).  The dashed line shows the position
of $-c$.}
\end{figure}

Moving from the forward to the reverse bias regime, the structure of the Bethe roots changes.  From a forward
bias root distribution (Figures \ref{fig:roots_forward_full}, \ref{fig:roots_forward}), increasing $-c$ pushes
out a complex conjugate pair of roots along with the single root on the positive real axis (Figure
\ref{fig:roots_interm}).  As $-c$ increases further, the complex conjugate pair pinch in to the real axis until some
point where they ``collapse'' (Figure \ref{fig:roots_reverse}) -- one root returns to close the contour on the
positive real axis leaving an isolated pair of real roots at
\begin{equation}
   z^\pm = -c \pm e^{-\nu^\pm L}, \quad \nu^\pm > 0.
\end{equation}
In fact, there is a pair of isolated roots for all $k \in \mathbb{N}$ such that $-q^k c$ falls outside the contour of complex roots. 

For $L$ large, the contour crosses the real axis at \cite{dGE08}
\begin{equation}
  z^*(a) = \frac{1+4a+a^2-\sqrt{(1+4a+a^2)^2-4a^2}}{2a} \simeq \frac{1}{a},
\label{eq:zstar}
\end{equation}
where the approximation is for $a$ large. Therefore, the integer $m$ defined by $a$ and $c$ through the inequalities
\begin{equation}
    -q^m c < \frac{1}{a} < -q^{m-1} c,
\label{eq:mdef}
\end{equation}
is equal to the maximum number of pairs of real roots as $L \to \infty$. While we take our parameters to have generic values, it would be of interest to analyse the system at the resonant values of $a$, $c$ where the inequalities become equalities, see e.g. \cite{CrampeRS10}.
\begin{figure}[hb]
\centering
\subfigure{
  \includegraphics{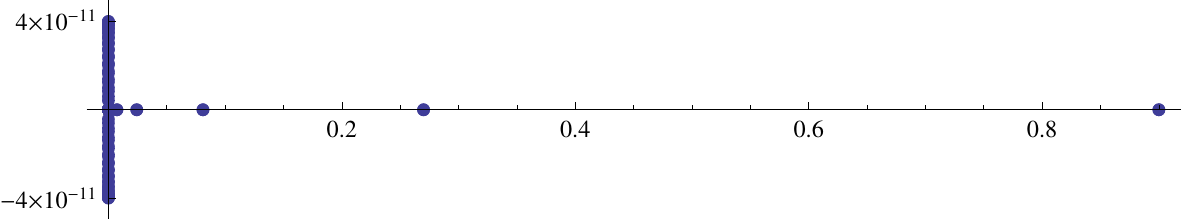}
 }
 \subfigure{
  \includegraphics{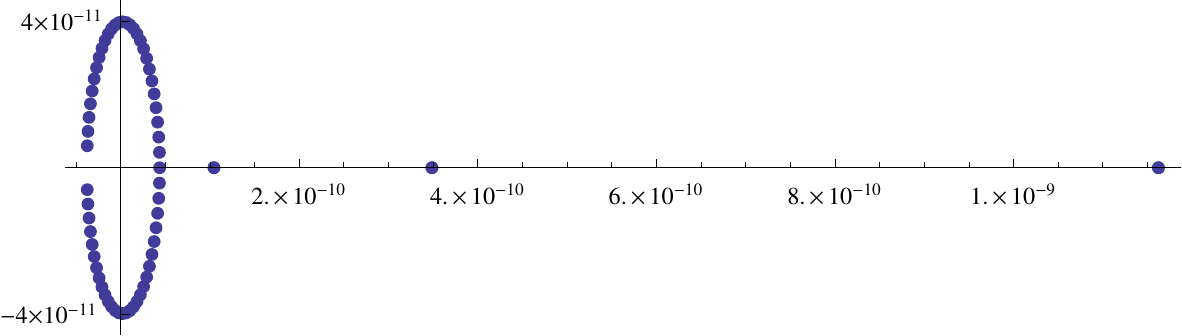}
 }

\caption{Complete root distribution (top) and zoomed in on the contour and first few pairs of isolated roots
   (bottom) for $L=100$, $\bar{a} = 3.09$, $c = -0.9$, $q = 0.3$ and $m=20$ (at this resolution only five real pairs are visible in the top figure).}
\label{fig:exsolutions}
\end{figure}
We define
\begin{equation}
   \bar{a} = q^{m-1} a,
   \label{eq:abar}
\end{equation}
then, for example, we will reach a maximum $m = 20$ pairs of real roots with the parameters
\begin{equation}
   \bar{a} = 3.09, \quad c = -0.9, \quad q = 0.3.
   \label{eq:exparams}
\end{equation}
The corresponding root distribution is displayed in Figure~\ref{fig:exsolutions}.

\begin{figure}[here!]
\begin{center}
   \includegraphics{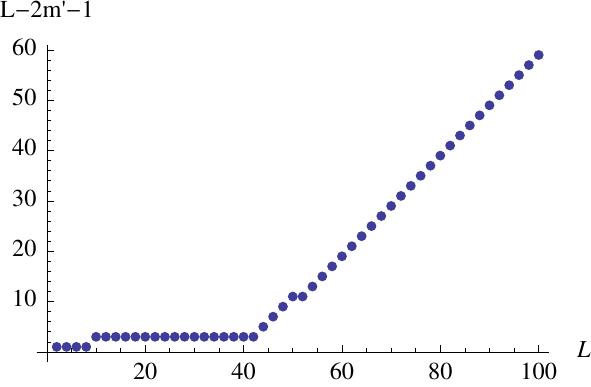}
   \caption{Number of complex roots in the contour ($L - 2m' - 1$) against $L$.  Only beyond $L = 40$ does the contour of complex roots begin to form.}
   \label{fig:Lbar}
\end{center}
\end{figure}

We denote by $m'$ the number of pairs of real roots for a particular value of $L$, with $m' \le m$. Then the number of roots on the contour is $L - 2m' - 1$, which is plotted in
Figure \ref{fig:Lbar} for the parameters \eref{eq:exparams}. There are three regions to consider:
\begin{enumerate}
   \item Isolated roots region: $L - 2m' - 1$ small and almost all roots occur as isolated pairs;
   \item Crossover region: The contour of complex roots begins to form, $m' \simeq m$;
   \item Asymptotic region: The number of roots on the contour, $L-2m'-1$ is large.
\end{enumerate}

%
%

\subsection{Isolated roots region}
\label{sec:allreal}

We first consider the case where $L - 2m' -1 $ is small and assume all roots are real, a situation similar to that encountered in the recent calculation of the PASEP current fluctuations \cite{dGE2011}. The $m'$ pairs of isolated roots are
\begin{equation}
    z_k^\pm = -q^k c \pm e^{-\nu_k^\pm L}\qquad k=0,\ldots,m'-1,
\end{equation}
and the remaing root is close to $-q^m c$.  Substituting these roots into the 
expression for the eigenvalue \eref{eq:pasep_en} and dropping exponentially small parts, the sum telescopes, leaving
\begin{equation}
    \mathcal{E}_1^{\rm{(is)}}  = 
    -(1-q) \frac{2q^{m-1}}{q^{m-1} + \bar{a}}
    -(1-q)q^{m'-1}\left(
              \frac{-qc}{1+q^{m'}c}+\frac{-q^2c}{1+q^{m'+1}c}
     \right).
\label{eq:E_CL_RR}
\end{equation}
The $L$ dependence enters through $m'$ and implies that the eigenvalue decays
as $q^{L/2}$ to a constant, which is $\mathcal{O}\left(q^m\right)$.  This
differs markedly from the asymptotic expression in the forward bias regime
\eref{eq:E_CL}.

\subsection{Asymptotic region}
\label{sec:revasymp}

We now analyse the third region, where the number of roots on the contour is large.  As we are interested in the asymptotic behaviour, we will assume we have reached the maximum number of pairs of real roots ($m' = m$).
We label the roots on the contour
\bea
   z_j, \quad j = 1 \ldots L - 2m - 1,
\eea
and the real roots, as before, as
\bea
    z_k^{\pm} = -q^k c \pm \e^{-\nu_k^\pm L}, \qquad \nu_k^\pm > 0,\qquad k = 0, \ldots, m-1.
\eea

The choice of branch cuts in the counting function \eref{eq:logtasepBAE}, \eref{eq:kernelDefFB} is suitable for the forward bias regime. In the reverse bias regime we must redefine the counting function. The form we use is
\begin{eqnarray}
  \i Y_L(z)& = & g(z) + \frac{1}{L} g_{\rm b}(z)
  + \frac{1}{L} \sum_{l=0}^{m-1} \left( K_{\rm{r}}(z_l^-,z) + K_{\rm{r}}(z_l^+,z)\right) \nonumber \\
&& \hspace{2.5cm} + \frac{1}{L} \sum_{l=1}^{L-2m-1} K(z_l,z) 
\label{eq:logtasepRB}
\end{eqnarray}
where now
\bea
   g(z) &=& \ln \left( \frac{z(1-qz)^2}{(1-z)^2}\right),     
   \nonumber\\
   g_{\rm b}(z) &=& \ln \left(\frac{1-q^2z^2}{z(1-z^2)}\right)
      + 2 \ln\left(
           c \frac {q^{m-1}z+\bar{a}}{q^{m-1}+q\bar{a}z}\frac{1+z/c}{1+qcz}
        \right), 
        \nonumber\\
  K(w,z) &=& -\ln \left( \frac{1 - qz/w}{1-qw/z} \frac{1-q^2zw}{1-zw}\right)
                        - \ln w, \label{eq:KernelDef} \\
  K_{\rm{r}}(w,z) &=& -\ln \left( \frac{1 - q^{-1}w/z}{1-qw/z} \frac{1-q^2wz}{1-wz} \right)
                        - \ln(qz). 
                        \nonumber
\eea
With this definition, the complex roots satisfy
\begin{equation}
  Y_L(z_j) = \frac{2\pi}{L} I_j, \quad j = 1, \ldots, L - 2 m - 1,
  \label{eq:Z=Icomplex}
\end{equation}
with the integers $I_j$ that correspond to the first excited state given by
\begin{equation}
  I_j = -\left(\frac{L}{2} - m\right) + j.
\end{equation}
The real roots satisfy
\begin{equation}
  Y_L(z_k^\pm) = \frac{2\pi}{L} I_k^\pm, \qquad k = 0, \ldots, m - 1,
\end{equation}
but the $I_k^\pm$ are not distinct.  To find numerical solutions, we use an alternative form of the counting
function that numbers all roots consecutively (see \ref{ap:numsols}).  However, the form \eref{eq:logtasepRB},
\eref{eq:KernelDef} is better suited to the following asymptotic analysis.

\subsubsection{Counting function for the complex roots}

We can account for the contribution of the isolated roots in \eref{eq:Z=Icomplex} and obtain Bethe equations for the complex roots alone. Ignoring exponentially small parts, the sum over real roots in \eref{eq:logtasepRB} telescopes,
\bea \label{eq:greal}
  \nonumber  &&   \sum_{l=0}^{m-1}\Bigg(K_r(z_l^-,z) + K_r(z_l^+,z) \Bigg) \\
  &\simeq& -2m \ln(q z)
          + 2\ln\left[
              \left(1+\frac{z}{q^m c}\right)
              \left(1+\frac{z}{q^{m-1}c}\right)
              \left(c + q z\right)
           \right] \nonumber\\
  && \hspace{0.5cm}
          -\ln(z + c)
          -\ln\left[
             \frac{(1+q^{m+1}c z)(1+q^m c z)}{(1+q c z)(1 + c z)}
           \right].
\eea

The complex roots lie on a simple curve in the complex plane, which approaches a closed contour as
$L\rightarrow \infty$.  From \eref{eq:zstar} and \eref{eq:abar}, we see that the size of the
contour scales as $q^{m-1}$.  We therefore change to the scaled variables
\begin{equation}
   z = q^{m-1} \zeta,
\end{equation}
and take into account the $2m$ real roots ``missing'' from the contour by defining a new counting function
\begin{equation}
    \overline{Y}_{L-2m}(\zeta) = \frac{L}{L-2m} Y_L(q^{m-1}\zeta).
\end{equation}
With the change to the scaled variables $\zeta$, we may drop  any $\mathcal{O}(q^m)$ contributions.
For example,
\bea
   g(z) = g(q^{m-1}\zeta)
        &=& (m-1)\ln q + \ln \left( \frac{\zeta(1-q^{m}\zeta)^2}{(1-q^{m-1}\zeta)^2}\right) \nonumber\\
       &\simeq& (m-1)\ln q + \ln \zeta.
\eea

Collecting terms by order of $1/(L-2m)$, the new counting function is given by
\begin{equation}
  \i \overline{Y}_{L-2m}(\zeta) = \bar{g}(\zeta) + \frac{1}{L-2m} \bar{g}_{\rm b}(\zeta)
       + \frac{1}{L-2m} \sum_{l=1}^{L-2m-1} \overline{K}(\zeta_l,\zeta),
\label{eq:cf}
\end{equation}
with
\bea
  \bar{g}(\zeta) &= \ln \zeta \label{eq:gSc} 
\nonumber \\
  \bar{g}_{\rm b}(\zeta) &= -\ln(\zeta)
         + 2\ln \left(\frac{\bar{a}\zeta}{1+q\bar{a}\zeta}\right) 
         + 2 \ln\left(\zeta+qc\right) + 2\ln\left(1 + \frac{\zeta}{c}\right), \label{eq:gbSc} \\
  \overline{K}(\omega, \zeta) &=
           -\ln \left(\frac{1- q\zeta / \omega}{1 - q\omega /\zeta} \right) -\ln(\omega),
\nonumber
\eea
where we recall that $\bar{a}$ is defined in \eref{eq:abar}.  The telescoped sum \eref{eq:greal} contributes terms to both $\bar g$ and $\bar{g}_{\rm b}$.
The Bethe ansatz equations for the scaled complex roots are
\begin{equation}
   \overline{Y}_{L-2m}(\zeta_j) = \frac{2\pi}{L-2m}I_j, \quad j=1,\dots,L-2m-1,
\label{eq:logtasepSc}
\end{equation}
with
\begin{equation}
   I_j=-\frac{L-2m}{2}+j.
\end{equation}
The eigenvalue, in terms of the scaled roots $\zeta_j$ is
\begin{equation}
    \mathcal{E}_1^{\textrm{(rev)}} = -\mathcal{E}_0^{(\textrm{rev})}
                   -\sum_{l = 1}^{L-2m-1} \bar{\varepsilon}(\zeta_l),
\label{eq:pasep_enSc}
\end{equation}
where
\begin{equation}
   \mathcal{E}_0^{(\textrm{rev})} = -2(1-q)q^{m-1}\left(
                           \frac{1}{q^{m-1}+\bar{a}} - \frac{qc}{1 + q^m c} \right), 
\label{eq:EzeroSc}
\end{equation}
includes $\mathcal{E}_0$ and the contribution from the real roots, and the scaled bare energy is
\begin{equation}
   \bar{\varepsilon}(\zeta) = -(1-q)^2q^{m-1}
            \frac{\zeta}{\left(1-q^{m-1}\zeta\right)\left(1-q^m\zeta\right)}.
\label{eq:epsdefSc}
\end{equation}

\section{Asymptotic Analysis}
The equations \eref{eq:logtasepSc} are logarithmic Bethe equations for the complex roots in the reverse bias
regime.  These equations have the same form as the forward bias equations \eref{eq:logtasepBAE} with the key
difference that there are now $L-2m-1$ roots instead of $L-1$ (recall that $m$ is defined by the condition \eref{eq:mdef}).  The method used in \cite{dGE08} to derive exact large $L$ asymptotics for the eigenvalue $\mathcal{E}_1$ can be applied to the reverse bias regime.  But we
require now that $L - 2m$ is large.

\begin{figure}[here!]
\begin{center}
\resizebox{6cm}{!}{\includegraphics{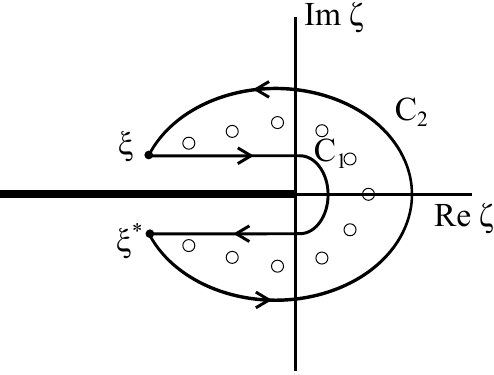}}
\caption{Integration contour enclosing the complex roots.
\label{contour2}}
\end{center}
\end{figure}

As a simple consequence of the residue theorem we can write
\bea\label{eq:sti}
   \frac{1}{L-2m}\sum_{j=1}^{L-2m-1}f(\zeta_i)\nonumber
   &=& \\
   &&\hspace{-2cm}\oint_{C_1+C_2} \frac{d\zeta}{4\pi \i} f(\zeta)\overline{Y}^{\prime}_{L-2m}(\zeta)
                          \cot \left( \frac{1}{2}(L-2m)\overline{Y}_{L-2m}(\zeta)\right),
\eea
where $C = C_1 + C_2$ is a contour enclosing all the complex roots $\zeta_j$ (see Figure \ref{contour2}) and
$f(\zeta)$ is a function analytic inside $C$.  We will use \eref{eq:sti} to write
an integro-differential equation for the counting function $\overline{Y}_{L-2m}(\zeta)$, but first we note some
properties of $\overline{Y}_{L-2m}(\zeta)$.

The complex roots $\zeta_j$ are connected by the contour $\textrm{Im}\left[\overline{Y}_{L-2m}(\zeta)\right] =
0$.  This contour extends to a point on the negative real axis, $\zetac$.  We fix $\xi$ and $\xi^*$,
the endpoints of $C_1$ and $C_2$, on this contour by
\begin{equation}
   \overline{Y}_{L-2m}(\xi^*) = -\pi + \frac{\pi}{L-2m}, \quad
   \overline{Y}_{L-2m}(\xi) = \pi - \frac{\pi}{L-2m}.
\label{eq:xidef}
\end{equation}
On $C_1$, inside the contour connecting the roots, the imaginary part of $\overline{Y}_{L-2m}(\zeta)$ is
positive; on $C_2$  it is negative.

Using \eref{eq:sti} in \eref{eq:cf}, we obtain a non-linear integro-differential
equation for $\overline{Y}_{L-2m}(\zeta)$,  which, using the properties just described, is written as an
integral over the contour of roots from $\xi^*$ to $\xi$, and correction terms integrated over $C_1$ and $C_2$:
\begin{eqnarray}
   \i\,\overline{Y}_{L-2m}(\zeta) &=& \bar{g}(\zeta) + \frac{1}{L-2m} \bar{g}_{\rm b}(\zeta)
       + \frac{1}{2\pi}\int_{\xi^*}^{\xi}
                \overline{K}(\omega,\zeta) \overline{Y}'_{L-2m}(\omega) \d \omega \nonumber \\ 
     && \hspace{-1cm} + \frac{1}{2\pi}\int_{C_1}
          \frac{\overline{K}(\omega,\zeta)\overline{Y}'_{L-2m}(\omega)}{1-\e^{-\i (L-2m) \overline{Y}_{L-2m}(\omega)}}\, \d \omega
       \nonumber \\
     && \hspace{-1cm} + \frac{1}{2\pi}\int_{C_2}
          \frac{\overline{K}(\omega,\zeta)\overline{Y}'_{L-2m}(\omega)}{\e^{\i (L-2m) \overline{Y}_{L-2m}(\omega)}-1}\,\d \omega.
\label{eq:intY}
\end{eqnarray}

The correction terms can be approximated as in \cite{dGE08} giving
\begin{eqnarray}
\i\,\overline{Y}_{L-2m}(\zeta) &=& \bar{g}(\zeta) + \frac{1}{L-2m} \bar{g}_{\rm b}(\zeta)
     +\frac{1}{2\pi}\int_{\zetac^-}^{\zetac^+}
                      \overline{K}(\omega, \zeta) \overline{Y}'_{L-2m}(\omega) \d \omega \nonumber\\
  &&{}\hspace{-1.8cm} + \frac1{2\pi} \int_{\xi^*}^{\zetac^-} \overline{K}(\omega,\zeta) \overline{Y}'_{L-2m}(\omega) \d \omega
  + \frac1{2\pi} \int_{\zetac^+}^{\xi} \overline{K}(\omega,\zeta) \overline{Y}'_{L-2m}(\omega) \d \omega \nonumber\\
  &&\hspace{-1.8cm} {} + \frac{\pi}{12(L-2m)^2} \left(
          \frac{\overline{K}'(\xi^*,\zeta)}{\overline{Y}'_{L-2m}(\xi^*)}
          - \frac{\overline{K}'(\xi,\zeta)}{\overline{Y}'_{L-2m}(\xi)}
   \right) + \mathcal{O}\left(\frac{1}{(L-2m)^4}\right).
\label{eq:intY_M}
\end{eqnarray}
Here, the derivatives of $\overline{K}$ are with respect to the first argument, and we have extended the
integration contour beyond the endpoints, so that it pinches the negative real axis at points
$\zetac^\pm=\zetac\pm \i 0$.

Equation \eref{eq:intY_M} is solved as follows. We Taylor expand the integrands of the integrals from $\xi^*$ to $\zeta_c^-$ and from $\zetac^+$ to $\xi$. Then we substitute the expansions
\begin{equation}
   \overline{Y}_{L-2m}(\zeta)=\sum_{n=0}^\infty (L-2m)^{-n}y_n(\zeta),
   \quad
   \xi=\zeta_c+\sum_{n=1}^\infty (L-2m)^{-n}(\delta_n + i\eta_n)
   \label{eq:expansion}
\end{equation}
and expand in inverse powers of $L-2m$. This yields a hierarchy of integro-differential equations for the functions $y_n(\zeta)$
\begin{equation}
   y_n(\zeta) = g_n(\zeta) + \frac{1}{2\pi\i} \int_{\zeta_{\rm c}^-}^{\zeta_{\rm c}^+}
                                                        \overline{K}(\omega,\zeta) y'_n(\omega) \,d\omega.
\label{eq:yn}
\end{equation}
The integral is along the closed contour following the locus of roots.  The first few driving terms
$g_n(\zeta)$ are given by
\begin{equation}
\renewcommand{\arraystretch}{1.2}
\begin{array}{r@{\hspace{2pt}}c@{\hspace{3pt}}l}
   g_0(\zeta) &=& -\i \bar{g}(\zeta), \\
   g_1(\zeta) &=& -\i \bar{g}_{\textrm b}(\zeta) + \kappa_1 + \lambda_1
   \widetilde{\overline{K}}(\zetac, \zeta), \\
   g_2(\zeta) &=&  \kappa_2 + \lambda_2 \widetilde{\overline{K}}(\zetac, \zeta) + \mu_2
   \overline{K}'(\zetac, \zeta), \\
   g_3(\zeta) &=&  \kappa_3 + \lambda_3 \widetilde{\overline{K}}(\zetac, \zeta) + \mu_3
   \overline{K}'(\zetac, \zeta)
                          + \nu_3 \overline{K}''(\zetac, \zeta). 
\end{array}
\label{eq:gs}
\end{equation}
The functions $\bar{g}(\zeta)$ and $\bar{g_{\textrm b}}(\zeta)$ are defined in \eref{eq:gSc}
and \eref{eq:gbSc}, and
\begin{equation}
   \widetilde{\overline{K}}(\zetac,\zeta) = -\ln(-\zetac) + \ln\left( \frac{1-q \zetac/\zeta}{1-q\zeta/\zetac} \right)
\end{equation}
arises when we evaluate $\overline{K}(\zetac^\pm, \zeta)$ and take the limit $\zetac^\pm \to
\zetac$.

The coefficients $\kappa_n$, $\lambda_n$, $\mu_n$ and $\nu_n$ are given in terms of
$\delta_n$, $\nu_n$, defined by \eref{eq:expansion}, and by derivatives of $y_n$ evaluated at
$\zetac$.  Explicit expressions are given in \ref{ap:Mcoef}.  We note that although the
definition of the counting function has changed, and despite expanding in $L-2m$ instead of $L$, the form
of these coefficients and the driving terms are unchanged from \cite{dGE08}.

\subsection{Solving for the $y_n$}
We may now solve \eref{eq:yn} order by order.  Working with the scaled roots, with the restrictions
on parameters of the reverse bias regime we can assume that $-q c$ lies inside the contour of
integration, and that $-\frac{1}{q\bar{a}}$ and $-c$ lie outside it.  The calculation follows in
the same way as in \cite{dGE08}, so we give here only the result for the counting function:
\bea
  \overline{Y}_{L-2m}(\zeta) &=& y_0(\zeta) + \frac{1}{L-2m} y_1(\zeta) + \frac{1}{(L-2m)^2} y_2(\zeta)
      \nonumber \\
          &&+ \frac{1}{(L-2m)^3} y_3(\zeta)
           + \mathcal{O}\left(\frac{1}{(L-2m)^4}\right)\label{eq:cfsol},
\eea
where
\bea
   y_0(\zeta) &=& -\i \ln\left(-\frac{\zeta}{\zetac}\right) \\
    \nonumber y_1(\zeta) &=& -\i \ln\left(-\frac{\zeta}{\zetac}\right)
         -2\i\ln\left(\bar{a}\right) -\lambda_1\ln(-\zetac) + \kappa_1 \\
     \nonumber
          && -2\i \ln\left[\frac{\qpch{-q c/\zeta}}{\qpch{-qc/\zetac}}\right]
          +\lambda_1 \ln\left[ \frac{\qpch{q\zetac /\zeta}}{\qpch{q \zeta/\zetac}} \right] \\
          &&- 2\i\ln\left[ \frac{\qpch{-\zeta/c}}{\qpch{-\zetac/c}} \right] \\
   \nonumber y_2(\zeta) &=& \kappa_2 -\lambda_2\ln(-\zetac)-\frac{\mu_2}{\zetac}
      + \lambda_2\ln\left[\frac{\qpch{q\zetac/\zeta}}{\qpch{q\zeta/\zetac}}\right]
      + \mu_2 \Psi_1(\zeta|q) \\
   \nonumber y_3(\zeta) &=& \kappa_3 -\lambda_3\ln(-\zetac)-\frac{\mu_3}{\zetac}
      + \frac{\nu_3}{\zetac^2}
      + \lambda_3\ln\left[\frac{\qpch{q\zetac/\zeta}}{\qpch{q\zeta/\zetac}}\right] \\
       && + \mu_3 \Psi_1(\zeta|q) - \nu_3 \Psi_2(\zeta|q) 
\eea

Here $\qpch{a}$ denotes the q-Pochhammer symbol 
\begin{equation}
   \qpch{a} = \prod_{k=0}^{\infty} (1-aq^k),
\label{pochh}
\end{equation}
and we have defined functions\begin{equation}
   \Psi_k(\zeta|q) =
        \psi_k(\zeta|q^{-1})-\psi_k(\zetac|q^{-1})
        -\left(\psi_k(\zeta|q) - \psi_k(\zetac|q)\right)
\end{equation}
with\footnote{
    The definition of $\psi_k(\zeta|q)$ used in \cite{dGE08} is incorrect for expressions such as
    $\psi_k(\zeta^{-1}|q^{-1})$.
}
\begin{equation}
   \psi_k(\zeta|q) = \sum_{n=0}^{\infty}\frac{1}{(\zetac-q^{n+1} \zeta)^k}\ .
\end{equation}

\subsection{Boundary conditions}
Substituting the expansions \eref{eq:expansion} into the boundary conditions
\eref{eq:xidef}, which fix the endpoints $\xi$ and $\xi^*$, we obtain
a hierarchy of conditions for $y_n(\zetac)$, e.g. 
\begin{eqnarray}
   \overline{Y}_{L-2m}(\xi)&=&y_0(\xi)+\frac{1}{L-2m}y_1(\xi)+\frac{1}{(L-2m)^2}y_2(\xi)+\ldots\nn
   &=& y_0(\zetac)+\frac{1}{L-2m}\left[y_1(\zetac)+y_0'(\zetac)(\delta_1+\i\eta_1)\right]+\ldots\nn
   &=& \pi-\frac{\pi}{L-2m}\ .
\label{eq:yexpand}
\end{eqnarray}
Solving this equation order by order, we find
\begin{equation}
    \lambda_1 = 2 \i, \qquad \zetac = -\frac{1}{\bar{a}},
\end{equation}
and
\begin{equation}
\renewcommand{\arraystretch}{1.2}
    \begin{array}{r@{\hspace{2pt}}c@{\hspace{3pt}}l}
    \lambda_3 &=& \mu_2 = \lambda_2 = \kappa_1= \kappa_2 = 0,\\
    \nu_3 &=& \dps \zetac \mu_3 = -\i \pi^2 \zetac.
    \end{array}
    \label{eq:constants}
\end{equation}

\subsection{Eigenvalue with largest non-zero real part}
As was done for the counting function, we use \eref{eq:sti} in \eref{eq:pasep_enSc} to obtain an
integro-differential equation for the eigenvalue.  Evaluating the resulting integrals in the same way as for the counting function itself, we obtain
\begin{equation}
   \mathcal{E} = -\mathcal{E}_0^{\textrm{(rev)}} 
           -\frac{L-2m}{2\pi} \oint_{\zetac} \bar{\varepsilon}(\zeta)\overline{Y}_{L-2m}'(\zeta)\, d\zeta
           -\i \sum_{n \geq 0} e_n (L-2m)^{-n},
\label{eq:energyMI}
\end{equation}
where the integral is over the closed contour on which the roots lie,
$\mathcal{E}_0^{\textrm{(rev)}}$ and $\bar \varepsilon$ are
given in \eref{eq:EzeroSc} and \eref{eq:epsdefSc}, and
\bea
   \nonumber e_0 &=& \lambda_1 \bar{\varepsilon}(\zetac),\nonumber \\
             e_1 &=& \lambda_2 \bar{\varepsilon}(\zetac) + \mu_2 \bar{\varepsilon}'(\zetac),\\
   \nonumber e_2 &=& \lambda_3 \bar{\varepsilon}(\zetac) + \mu_3 \bar{\varepsilon}'(\zetac)
                                                         + \nu_3 \bar{\varepsilon}''(\zetac).
\eea
Substituting the expansion \eref{eq:cfsol} for $\overline{Y}_{L-2m}(\zeta)$ into \eref{eq:energyMI} we arrive at the following result for the eigenvalue of the transition matrix with the largest non-zero real part
\begin{equation}
     \mathcal{E}_1^{\textrm{(rev)}}
           = \frac{q^{m-1}}{(L-2m)^2} \frac{\pi^2(1-q) \bar{a}}{q^{2(m-1)} - \bar{a}^2}
          + \mathcal{O}\left(\frac{1}{(L-2m)^3}\right).
\label{eq:E_CL_RB}
\end{equation}
Making the substitution $\bar{a} = q^{m-1}a$, we see that this is
\begin{equation}
     \mathcal{E}_1^{\textrm{(rev)}}
           = \frac{1}{(L-2m)^2} \frac{\pi^2(1-q)}{a^{-1} - a}
          + \mathcal{O}\left(\frac{1}{(L-2m)^3}\right),
\end{equation}
which differs from the expression in the forward bias regime \eref{eq:E_CL} only by changing $L \to L -2m$. In Section~\ref{se:conc} we discuss the physical interpretation of this result.

\section{Crossover region}
We have found two very different expressions for the eigenvalue: $\mathcal{E}_1^{\textrm{(is)}}$ in \eref{eq:E_CL_RR} describing the
relaxation for relatively small $L$, dominated by the isolated real roots, and $\mathcal{E}_1^{\textrm{(rev)}}$ in \eref{eq:E_CL_RB} describing the asymptotic relaxation when $L - 2m$ is large.  As $L$ increases, the contour of complex roots begins to form.  The first
approximation is no longer valid, and the second cannot be applied until the contour is sufficiently large.

Nevertheless, we can compare the magnitude of the two expressions in this region.  Assuming $m' = m$, we can
approximate \eref{eq:E_CL_RR} as
\begin{equation}
    \mathcal{E}_1^{\textrm{(is)}}
        \simeq -q^{m-1}(1-q)\left( \frac{2}{\bar{a}} - (1 + q) qc \right).
\end{equation}

\begin{figure}[here!]
\begin{center}
\includegraphics{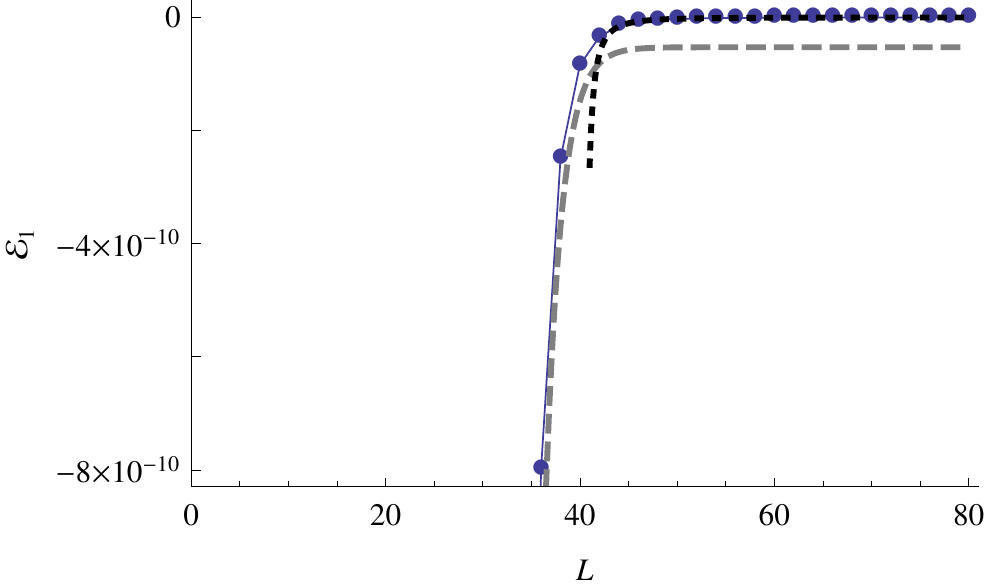}
\caption{$\mathcal{E}_1^{\textrm{(is)}}$ (dashed), $\mathcal{E}_1^{\textrm{(rev)}}$ (dotted), 
and numerical value (datapoints connected by thin solid line) for $m=20$ and parameters \eref{eq:exparams}.
\label{fig:E1_crossover}}
\end{center}
\end{figure}

The asymptotic expression \eref{eq:E_CL_RB} can be approximated as
\begin{equation}
    \mathcal{E}_1^{\textrm{(rev)}} \simeq
           -q^{m-1}(1-q)\frac{\pi^2}{(L-2m)^2} \frac{1}{\bar{a}}.
\end{equation}
The contour begins to form when $L-2m = 2$, and at this point we see that
the two expressions are comparable.  Figure \ref{fig:E1_crossover} shows the two expressions and
the numerically calculated value in the crossover region.

\section{Conclusion}
\label{se:conc}
We have computed the relaxation rate on the coexistence line of the partially asymmetric exclusion process
with reverse biased boundary rates. Such rates introduce a length scale $m \approx -\log(-ac)/\log q$ in the system, determined by \eref{eq:mdef}. There are two distinct sub regimes. When the system size is small compared to the reverse
bias, i.e. $L \lesssim 2m$, the relaxation is exponential in the system size and the inverse relaxation time
is given by \eref{eq:E_CL_RR}. For large systems, $L\gg 2m$, the relaxation on the coexistence line is
diffusive but the inverse relaxation time vanishes with the square of a reduced system size $L-2m$. This
behaviour  can be obtained by an effective one particle diffusion model as in \cite{GierE06,KolomeiskySKS} but
in a reduced system of size $L-2m$.

We thus observe the effects of the reverse bias on the relaxation rate even with non-zero (but small) rates in the forward direction -- provided the rates in the reverse direction are large enough.  We suspect that in this case the forward contribution to the stationary current of particles is not strong enough to destroy the reverse bias phase, contrary to the argument in \cite{PASEPstat2}. It would
be of interest to revisit the calculation of the stationary current to see whether this is indeed the case.

The corresponding physical interpretation of these results is the following. In the setup of this paper, where the bulk bias is from left to right, the system will fill up completely on the right hand side over a length $2m$. In the remaining space of size $L-2m$, the system behaves as a regular PASEP on the coexistence line, and forms a uniformly distributed domain wall. As far as we are aware, the stationary density profile of the PASEP in the reverse bias regime has not been computed analytically, and it would be interesting to see whether this picture is confirmed by such a calculation.

\section*{Acknowledgments}
Our warm thanks goes to Fabian Essler for discussions. Financial assistance from the Australian Research Council is gratefully acknowledged.

\appendix

\section{Numerical solutions of the Bethe equations \label{ap:numsols}}
The logarithmic form of the Bethe equations is numerically stable for large $L$, as long as an appropriate
form of the counting function is used.  In the forward bias regime, that form is \eref{eq:logtasepBAE}.  For
the reverse bias regime the situation is more complicated.  The form \eref{eq:logtasepRB} is suited to the
asymptotic analysis -- by assuming the positions of the isolated roots, we choose the branch cuts of the
counting function to number the complex roots consecutively.  However, to find numerical solutions, we
arrange for the counting function to number all roots consecutively.  
That is, we define a counting function $Y_{L,m'}$ so that the Bethe equations become
\begin{equation}
    Y_{L,m'}(z_j) = \frac{2\pi}{L} I_j, \quad j = 1,\ldots,L-1,
\end{equation}
with
\begin{equation}
    I_j = -L/2 + j.
\end{equation}
The complex contour is formed by the roots
\begin{equation}
    z_j, \qquad j = 1\ldots L/2-m'-1,\quad L/2+m',\ldots,L-1,
\end{equation}
while the isolated roots are
\begin{equation}
    z_j = z_k^- = -q^k c - e^{-\nu_k^-L},\quad j = L/2 - k - 1,
\end{equation}
and
\begin{equation}
    z_j = z_k^+ = -q^k c + e^{-\nu_k^+L},\quad j = L/2 + k,
\end{equation}
for $k = 0,\ldots m'-1$.

We can think of \eref{eq:logtasepRB}, with $m = 0$ (no isolated roots), as $Y_{L,m'=0}$, which then is modified as
$m'$ increases.  For example, the first pair of isolated roots are $z_0^\pm$.
We work directly with the $\nu_k^\pm$ to reduce the required floating
point precision, and for $z_0^+$ rewrite the boundary term
\begin{equation}
    \ln\left(z_0^+ + c\right) = \ln\left(\left(-c + e^{-\nu_0^+L}\right) + c\right) = -\nu_0^+L.
\end{equation}
For $z_0^-$ we must also consider the phase as
\begin{equation}
    \ln\left(z_0^- + c\right) = \ln\left(- e^{-\nu_0^-L}\right) = -\nu_0^-L \pm \i \pi.
\end{equation}
Referring to Figure \ref{fig:roots_interm}, $z_0^-$ comes from the pushed out complex root with negative
imaginary part, which tends to
\begin{equation}
    -c - e^{-\nu_0^-L} -\i 0^-.
\end{equation}
Therefore, the appropriate choice of phase is
\begin{equation}
    \ln\left(z_0^- + c\right) = -\nu_0^-L - \i \pi.
\end{equation}

The interaction terms from $K(w,z)$ must also be handled similarly.  Roots $z_k^\pm$ and $z_{k-1}^\pm$ differ
by a factor of $q$, apart from the exponentially small correction.  This gives terms such as
\begin{equation}
   \ln\left(\pm e^{-\nu_k^\pm L} \mp e^{-\nu_{k-1}^\pm L} \right).
\end{equation}
When the argument of the logarithm is negative,\footnote{Note that we assume that $\nu_k^\pm > \nu_{k'}^\pm$ for $k < k'$, based on the numerical solutions.} we must choose the appropriate phase. Again, we consider how the isolated roots form, but must now consider their phase relative to all other roots.

With this form of the counting function, we have found numerical solutions for large $L$ and $m'$ using
standard root finding techniques.  The key difficulty that remains is finding the transition point from a pair
of complex roots pinching the real axis (Figure \ref{fig:roots_interm}) to a pair of isolated real roots
(Figure \ref{fig:roots_reverse}).

\section{Expansion coefficients\label{ap:Mcoef}}
In this appendix we list the coefficients arising in the expansion
(\ref{eq:yn}), (\ref{eq:gs}) of the integral equation for the counting
function $\overline{Y}_{L-2m}(\zeta)$. In the following list we abbreviate $y'_n(\zetac)$
by $y'_n$. We note that by definition $\delta_n$ and $\eta_n$ are real quantities.
\bea
\kappa_1 &=& - y_0' \delta_1,\\
\lambda_1 &=&  y_0' \frac{\eta_1}{\pi},\\
\kappa_2 &=& -y_0' \delta_2 - y_1' \delta_1 -
  \frac12 y_0'' (\delta_1^2 - \eta_1^2), \\ 
\lambda_2 &=& \frac1{\pi}\left( y'_0 \eta_2 + y'_1\eta_1 + y_0'' \delta_1
\eta_1\right),\\ 
\mu_2 &=& y'_0 \frac{\delta_1\eta_1}{\pi},\\
\kappa_3 &=& -y_0'\delta_3 - y_1'\delta_2 -y_2'\delta_1 - y_0'' (\delta_1\delta_2 -
\eta_1\eta_2) -\frac12 y_1''(\delta_1^2-\eta_1^2) \nonumber\\
&& {} - \frac16 y_0'''\delta_1(\delta_1^2 - 3\eta_1^2), \\
\lambda_3 &=& \frac1\pi \left( y_0'\eta_3 + y_1'\eta_2 + y_2'\eta_1 +
y_0''(\delta_1\eta_2 +\delta_2\eta_1) + \delta_1\eta_1y_1''\right.
\nonumber\\
&&\left. {} +\frac16 y_0''' \eta_1 (3\delta_1^2 - \eta_1^2) \right),\\
\mu_3 &=& \frac1\pi \left[ y_0'(\delta_1\eta_2 + \delta_2\eta_1) +
y_1'\delta_1\eta_1 + y_0''\eta_1(\delta_1^2-\frac{\eta_1^2}{3})
+\frac{\pi^2y_0''}{6y_0'^2} \eta_1\right],\\
\nu_3 &=& \frac1{6\pi}\left( y_0'\eta_1(3\delta_1^2-\eta_1^2) - \pi^2 \frac{\eta_1}{y'_0}\right).
\eea

\section*{References}
\providecommand{\newblock}{}

\end{document}